\documentclass[conference]{IEEEtran}
\IEEEoverridecommandlockouts
\usepackage{cite}
\usepackage{amsmath,amssymb,amsfonts}
\usepackage{algorithmic}
\usepackage{graphicx}
\usepackage{textcomp}
\usepackage{xcolor}
\usepackage{subcaption}
\usepackage{hyperref}
\usepackage[nameinlink, capitalise]{cleveref}

\usepackage{circuitikz}

\usepackage{enumitem}

\def\BibTeX{{\rm B\kern-.05em{\sc i\kern-.025em b}\kern-.08em
    T\kern-.1667em\lower.7ex\hbox{E}\kern-.125emX}}
\begin{document}

\title{Advisory Tool for Managing Failure Cascades in Systems with Wind Power
\thanks{\it We thank MIT UROP, MITEI, and the NSF EAGER project \#2002570 for funding, and Dan Wu, Xinyu Wu, and Miroslav Kosanic for discussions.}
}

\author{\IEEEauthorblockN{Siyu Liu}
\IEEEauthorblockA{\textit{Massachusetts Institute of Technology}\\
Cambridge, MA, USA \\
\url{eliu24@mit.edu}}
\and
\IEEEauthorblockN{Marija Ili\'{c}, IEEE Life Fellow}
\IEEEauthorblockA{\textit{Massachusetts Institute of Technology} \\
Cambridge, MA, USA \\
\url{ilic@mit.edu}}
}

\maketitle

\begin{abstract}
This paper concerns the resilience of systems with wind power upon wind reduction by evaluating the potential of corrective actions, such as generation and load dispatch, on minimizing the effects of transmission line failures. Three functions (grid, consumer-centric loss, and resilience impact) are used to statistically evaluate the criticality of initial contingent failures and wind reductions. Our model is learned with Monte Carlo, convex optimization, and adaptive selection, illustrated on the IEEE-30  and IEEE-300 bus systems with both AC and DC models. We highlight the impact of wind reductions and propose physically implementable solutions.

\end{abstract}


\section{Introduction} \label{Secion 1: Introduction}


Modern power systems are prone to many unpredictable component failures. Past events have shown that large scale blackouts are typically results of sequential failures of transmission lines, called \textit{failure cascades.} Cascades tend to evolve quickly, leaving only up to 15 minutes for the system operators to take corrective actions before the failure propagates\cite{puerto_rico}. Wind power integration introduces additional unpredictability and potential congestion-induced failures. Due to the fast-evolving nature of wind power and failure propagation, it is particularly important to understand the cascade patterns and their relations to sudden wind reduction, in order to advice system operators during such extreme events. Failure cascades are extremely computationally expensive to analyze due to the large number of nonlinear relations. Most of the early work concerns analysis of cascades and holds a pessimistic outlook at wind penetration due to its unpredictability\cite{earlyworks}.

In this paper, we go beyond analysis and explore corrective actions to minimize the effects of equipment failures that may be exacerbated by wind penetration. This is done by using probabilistic methods for deriving statistical information about the most effective corrective actions, such as generation re-dispatch and preemptive load shed. These are needed to prevent cascading failures as the actual events occur in on-line operations; they can also be used to enhance today's industry manuals. Most of the studies concern  wind penetration risk by solving a DC power flow model, which under-estimates the effects of failures and does not always provide physically implementable solutions due to reactive power and voltage constraints \cite{Cetinay}, \cite{wind1}-\cite{wind4}. To overcome the high complexity of power flow analysis, researchers have also sought various statistical methods, including random graphs, branching processes, and flow dynamics models --- all typically embed non-trivial constraint relaxations \cite{song}-\cite{bg-zhang2}. However, very few studies consider physically-implementable generation re-dispatch \cite{correction1, wind1} or load shedding. Those that do are only done for the purpose of minimizing grid-centric cost \cite{sinha, r-n, ShiLiu}. Others, including \cite{Cetinay} and \cite{Wu}, by implement a constant factor load shed algorithm without any ``smart scheduling.'' To the best of our knowledge, there is so far no statistical model for load shed prediction and management. To remedy these shortcomings, we adopt a flow-free approach using the influence model (IM). The IM is a Markovian model that, given the network profile at each time step, computes the link failure and load shed probability at all links and buses. It is straightforward to construct and fast to implement, reducing the prediction tasks to matrix multiplication and completely eliminating the burden of flow computation. We borrow insights from \cite{Wu} and \cite{song} to train the model using Monte Carlo, convex optimization, and adaptive selection. Our work drastically augments the scope of previous studies in both methodology, results, and applicability. 

In this paper, \cref{Section 2: Influence Model} briefly reviews the IM, and \cref{Section 3: Data Generation} outlines our simulation process. \cref{Section 4: Corrective Actions} proposes three metrics to evaluate various corrective actions and underlines the significance of our findings for systems with wind power. \cref{Section 5: Prediction Accuracy} summarizes the  prediction accuracy and time complexity of the IM. \cref{Section 6: Large System} demonstrates a comparable effectiveness on a large-scale system.
Finally, \cref{Section 7: Advisory tool for operators} introduces its use as an  advisory tool to system operators. 

\section{The Influence Model for Loss Prediction} \label{Section 2: Influence Model}

The IM is a Markovian-like model whose dynamics are described by the state variable transitions. We propose two IM, first for link failure prediction through matrix $D,$ and second for load shed prediction through matrix $E.$ Both models operate on a $(N_{br}\times 1)$ network state vector $s$ that stores the status of all links in binary, where $N_{br}$ is the number of transmission lines (branches) in the network. Given the $i$-th link, $s_i[t]=1$ indicates that link $i$ is alive at time $t,$ and $s_i[t]=0$ indicates that it has failed. 

\subsection{Matrix D for link failure prediction} \label{Section 2.1: D Matrix}
The link failure IM predicts the subsequent network state $s[t+1]$ given the current state $s[t]$ and trained parameters, which we define as follows.

\begin{itemize}[leftmargin=*]
    \item Transition probability matrices $A^{01}$ and $A^{11},$ both of size $(N_{br}\times N_{br}),$ where
        \begin{gather} \label{eq:A_prob}
            A^{11}_{ji}:=\mathbb{P}(s_i[t+1]=1|s_j[t]=1),\\
            A^{01}_{ji}:=\mathbb{P}(s_i[t+1]=1|s_j[t]=0).
        \end{gather}
    \item The weighted influence matrix $D$ of size $(N_{br}\times N_{br}),$ where the entry $d_{ij}$ represents the proportional effect from the link $j$ to $i.$ It can be interpreted as the ratio of the influence from link $j$ to $i$ among all links over $i.$
    \item The bisection threshold vector $\epsilon$ of size $(N_{br}\times 1).$ We determine $\epsilon_i$ for each initial failure profile by examining the $s_i$ sequence in all samples.
\end{itemize}

$A^{01}$ and $A^{11}$ are obtained through a Monte Carlo experiment, and the $D$ entries are obtained by solving an optimization problem as outlined in \cite{Wu}.

The transition probabilities from $A^{11}$ and $A^{01}$ weighted by the influence factors in $D$ gives the prediction $\widetilde{s_i}[t+1]:$
\begin{equation} \label{eq:s[t+1]} \widetilde{s_i}[t+1] =  \sum_{j=1}^{N_{br}} d_{ji}\left(A^{11}_{ji}s_j[t]+A^{01}_{ji}(1-s_j[t])\right),
\end{equation}
where we predict link remaining healthy when  $\widetilde{s_i}[t+1]\geq \epsilon_i.$

\subsection{Matrix E for load shed prediction} \label{Section 2.2: E Matrix}
The link failure IM predicts the $(N\times 1)$ load binary vector $l[t]$, where $l_i[t]=1$ indicates full service, and $l_i[t]=0$ indicates load reduction. The prediction is done based on the network state $s[t]$ and trained parameters, defined as follows.

\begin{itemize}[leftmargin=*]
    \item Transition probability matrices $B^{01}$ and $B^{11}$ of sizes $(N_{br}\times N)$ that define the weighted influences from links to buses, where $N$ is the number of buses. $B^{01}$ and $B^{11}$ defined as:
        \begin{gather} \label{eq:B_prob}
            B^{11}_{ji}:=\mathbb{P}(l_i[t]=1|s_j[t]=1),\\
            B^{01}_{ji}:=\mathbb{P}(l_i[t]=1|s_j[t]=0).
        \end{gather}    
    \item The weighted influence matrix $E$ of size $(N\times N_{br})$ defines the weighted influences from links to buses. Each entry $e_{ij}$ denotes the proportional influence of link $j$ on bus $i.$
    \item The bisection threshold vector $\delta$ of size $(N\times 1).$ We determine $\delta_i$ for each initial contingency profile by examining all samples where load shed has occurred at bus $i.$
\end{itemize}

$B^{01}, B^{11},$ and $E$ are obtained through a Monte Carlo experiment and convex optimization with similar to \cite{Wu}.

The probability of the system being able to serve full load at bus $i$ is calculated by a weighted sum of the influence from all links, using the trained parameters in $B^{01},$ $B^{11},$ and $\delta_i:$
\begin{equation} \label{eq:l[t]} \widetilde{l_i}[t] =  \sum_{j=1}^{N_{br}}e_{ij}\left(B^{11}_{ji}s_j[t]+ B^{01}_{ji}(1-s_j[t])\right),
\end{equation}
where we predict full service when $\widetilde{l_i}[t] \geq \delta_i.$

The run time for building the IM is dominated by the optimization step to obtain $D$ and $E,$ which takes $O(N^2 N_{br}^2).$

\section{Sample Pool Generation} \label{Section 3: Data Generation}

\begin{figure}[h]
\centering
\includegraphics[width=0.3\textwidth]{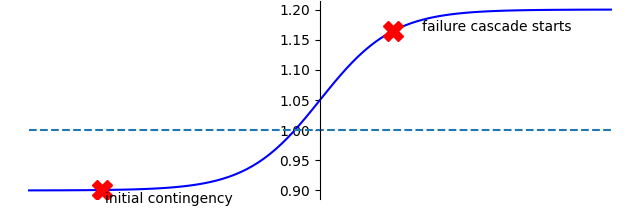}
\captionsetup{font={footnotesize}}\caption{Wind Reduction. Before wind power reduction, the net system load is at $\times .9$ base load. A $30\%$ wind reduction causes the net system load to rise to $\times 1.2$ base load, causing the second round of cascade failures.} \label{fig: wind_reduction}
\end{figure}

To examine the effects of sudden wind reduction, we first simulate the network under normal conditions. We introduce a random link failure under normal conditions where the loading level (base load - wind power) is nominal. If the failure does not lead to a complete blackout, we continue the simulation by introducing a wind reduction while the system is operating under $(N-2)$ conditions. The reduction level ranges from $10\%$ to $70\%$ of the base load, causing the net loading level to rise up to $\times 1.7$ the original loading level. We analyze the additional network congestion and failures caused by this load increase. \cref{fig: wind_reduction} provides a visualization to this process.

As there is no standard oracle for assessing failure cascades at present, we base our experiments on the CFS oracle proposed in \cite{CFS}. The CFS oracle is similar to the short-term OPA oracle, except that line outages are treated deterministically and it does not apply optimal re-dispatch during a failure cascade. Instead, it only sheds load or curtails generation if system-wide power mismatch occurs. For samples where no corrective actions are taken, we follow the CFS oracle exactly, and for sample where corrective actions are applied, we follow a relaxed version of the CFS oracle by allowing re-dispatch during the cascade. This relaxation is realistic, as the time between two failures can be as long as 15 minutes to allow the re-dispatch \cite{puerto_rico}. In all our experiments, we initialize the network as fully functional, randomly select two initial contingencies, and determine the cascade sequence following the oracle. Long term thermal condition is used when all links are fully functional, and it changes to short term thermal conditions once failures occur (which we assume to be $1.05\times$ long term). After each failure, we solve the DC/AC PF/OPF problem using the MATPOWER Toolbox \cite{MATPOWER}. Our three sets of experiments are defined with parameters as follows.

\textit{Experiment 1: No corrective action.}
In this experiment, we simply record the network status and loading levels at each bus at each step of the cascade without any corrective actions.

\textit{Experiment 2: Generation re-dispatch for full service.}
We re-dispatch generation whenever new link failures occur by solving for OPF under both uniform generation cost and bus-specific generation cost provided by \cite{MATPOWER}. We aim to serve all loads in full and only shed load uniformly in scale when unable to serve in full.

\textit{Experiment 3: Generation re-dispatch (smart scheduling).}
This experiment resembles \textit{Experiment 2}, except that instead of aiming for full service, we find the OPF solution that minimizes cost of shedding load, which we assume to be either uniform or priority-based. Notably, no links fail in this experiment, as the optimization step observes link constraints and maximum service as part of the optimization objective.

Our experiments are done on the IEEE-30 system with initial loading being $c$ times the test case from \cite{MATPOWER}. The $c$ ranges from $0.9$ to $1.8$ in $0.1$ increments. 

\section{Effect of Corrective Actions} \label{Section 4: Corrective Actions}

We examine the effects of corrective actions under unexpected wind reduction under DC and AC models. We propose two loss functions to quantify grid-centric and consumer-centric loss of each cascade sequence, as well as a resilience impact function to evaluate the loss given the base load and wind reduction in \cref{Section 4.1: grid loss} -- \cref{section 4.5: structures} analyze the effects of different corrective actions assessed by these functions and system-wide structural patterns.

\subsection{Grid-Centric Loss} \label{Section 4.1: grid loss} 
For each cascade sample, grid-centric loss is defined as
\begin{equation} \label{eq: link fail loss}
    G(p) = \sum_{b=1}^{N_{br}} C(b)\cdot e^{-0.2t_b},
\end{equation}
where $G(p)$ is the link failure loss for initial network profile $p,$ $C(b)$ the cost on branch $b,$ proportional to its maximum thermal capacity, and $t_b$ the life time of $b.$ The discounting factor $e^{-0.2t_b}$ is to penalize early failures.

In \textit{Experiments 1 \& 2}, links fail more frequently and earlier on in the cascade at higher initial loading levels. Even in the only two loading levels where \textit{Experiment 2} only successfully initializes, the loss of link failure is much greater than that in \textit{Experiment 1}. This demonstrates that PF models underestimates failure sizes. There is no observable difference between re-dispatching with actual or uniform generation cost in \textit{Experiments 2 \& 3}. \cref{fig: linkfail DC} illustrates these results.

\subsection{Consumer-Centric Loss} \label{Section 4.2: consumer loss} 

For each cascade sample, consumer-centric loss is defined with the formula as follows.

\begin{equation}\label{eq: load shed loss}
    L(p) = \sum_{l=1}^{N}\sum_{t=1}^{T_k-1} C(l) \cdot LS_l(t) e^{-0.2t},
\end{equation}
where  $L(p)$ is the load shed loss for initial network profile $p.$ $C(l)$ the load priority, and $LS_l(t)$ the amount of load shed between time steps $t$ and $t+1$ at bus $l.$ The expression is similarly time -- discounted by $e^{-0.2t}.$

\begin{figure}[hbtp]
\begin{minipage}{0.24\textwidth}
\includegraphics[width=1\textwidth]{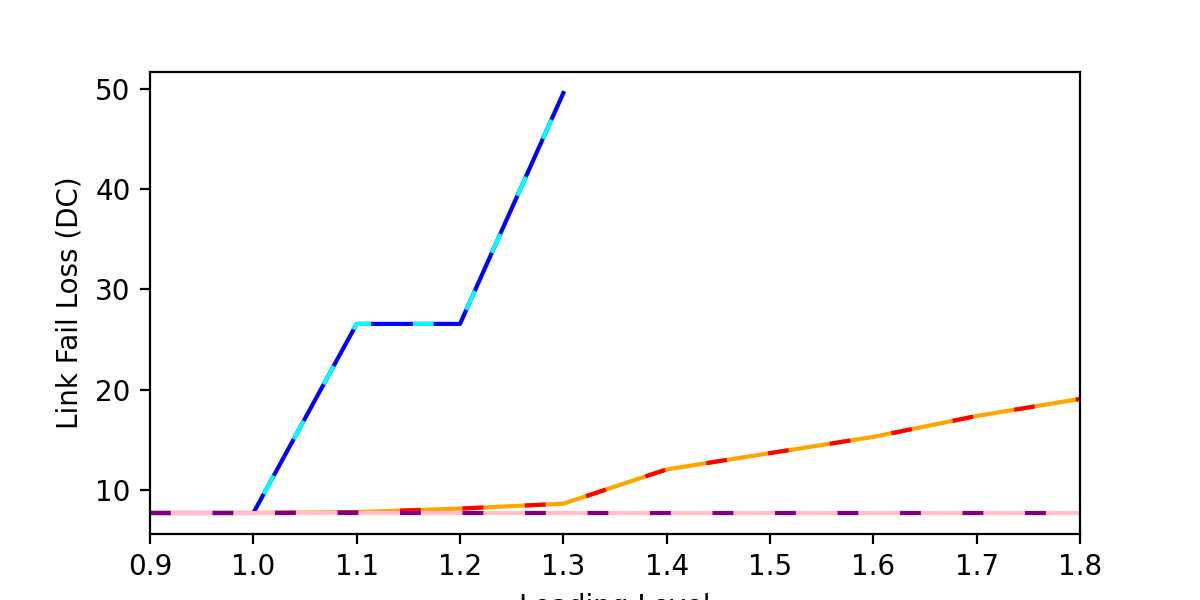}
\captionsetup{font={footnotesize}}\caption{Link Fail Loss (DC).} 
\label{fig: linkfail DC}
\end{minipage}\hfill
\begin{minipage}{0.24\textwidth}
\includegraphics[width=1\textwidth]{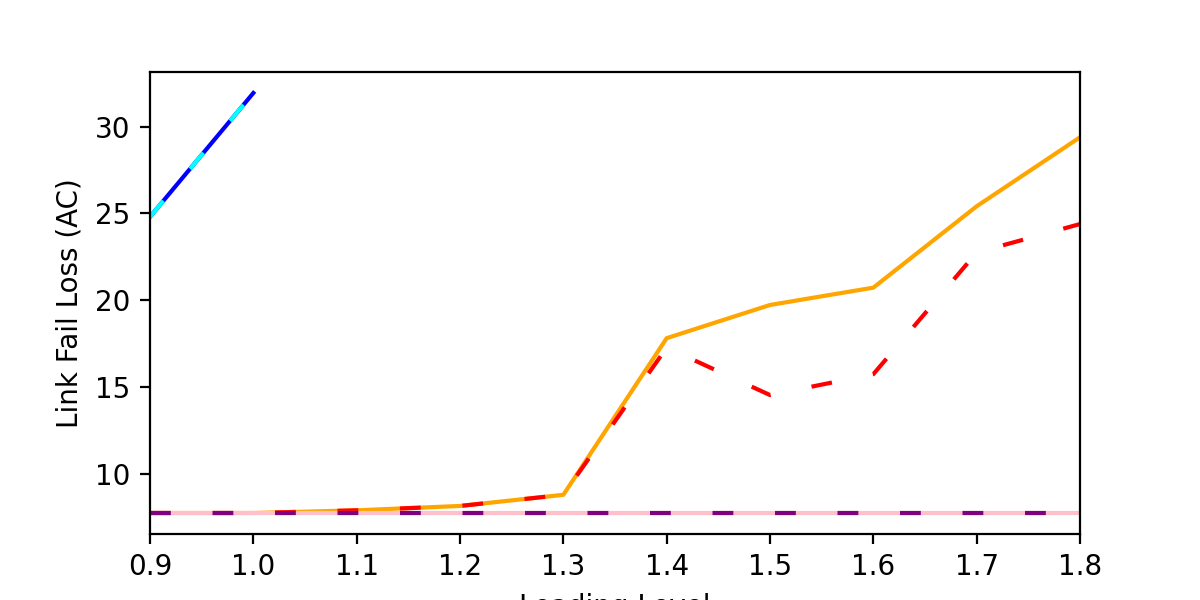}
\captionsetup{font={footnotesize}}\caption{Link Fail Loss (AC).}
\label{fig: linkfail AC}
\end{minipage}
\begin{minipage}{0.24\textwidth}
\includegraphics[width=1\textwidth]{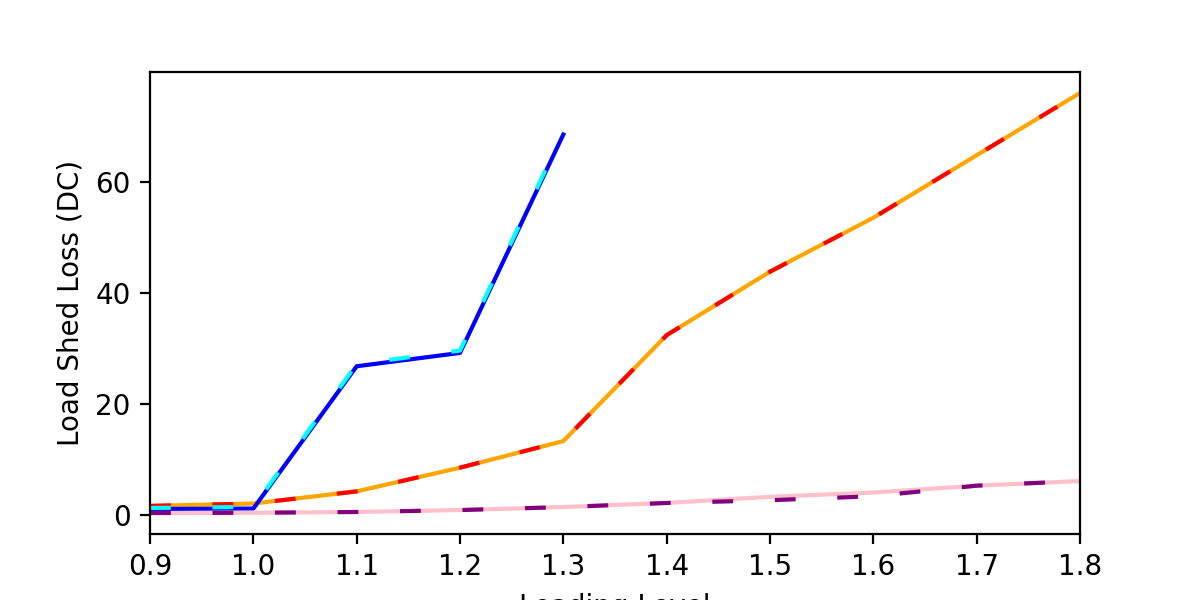}
\captionsetup{font={footnotesize}}\caption{Load Shed Loss (DC).} 
\label{fig: loadshed DC}
\end{minipage}\hfill
\begin{minipage}{0.24\textwidth}
\includegraphics[width=1\textwidth]{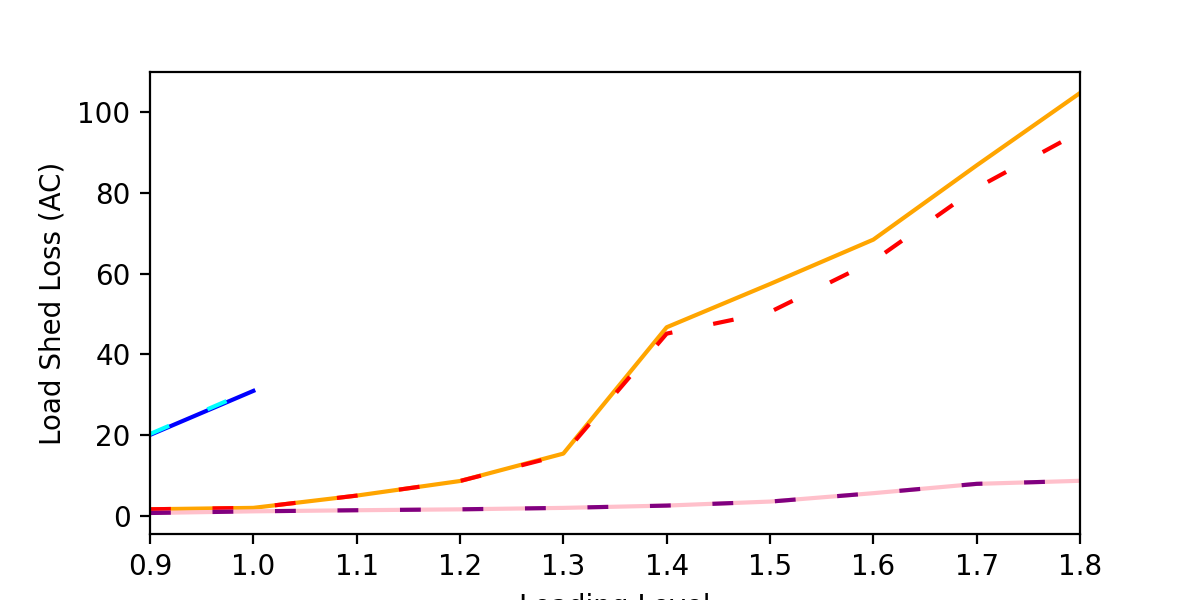}
\captionsetup{font={footnotesize}}\caption{Load Shed Loss (AC).} 
\label{fig: loadshed AC}
\end{minipage}
\centering
\begin{minipage}{0.3\textwidth}
\includegraphics[width=1\textwidth]{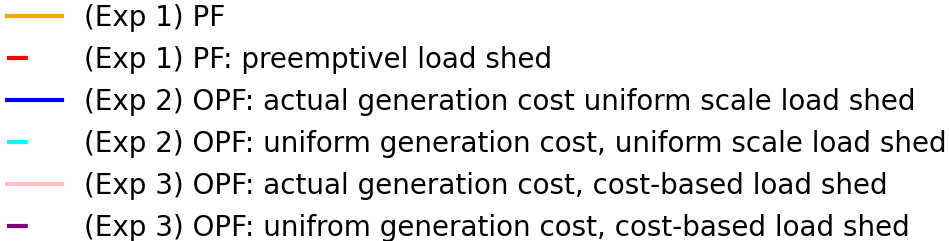}
\end{minipage}
\captionsetup{font={footnotesize}}\caption{Grid-centric and consumer-centric loss over under various corrective actions for DC and AC models.}
\label{fig:all plots}
\end{figure}

Our experiments yield one notable finding. If corrective actions are taken promptly, we may preserve infrastructure integrity in full without significant service reduction. In particular, load shed loss is minimized under \textit{Experiment 3}'s smart scheduling, when we run OPF with cost-based load shed. As such, the flow on all links are within their capacities and no cascade incurs. As illustrated in \cref{fig: loadshed DC}, comparing the load shed across all three experiments, \textit{Experiment 3} reduces the consumer-centric loss to much less than that in \textit{Experiments 1 \& 2}. The passive, emergency load shed in \textit{Experiment 2} incurs the greatest loss. Whether the cost of generation varies at different buses does not induce significant difference in the load shed loss. 

\subsection{Resilience Impact} \label{Section 4.3: resilience}

We propose the following equation to measure network resilience, where  $R(p, \Delta w)$ is network resilience for initial profile $p,$ wind reduction $\Delta w,$ and $p'$ the network profile during the wind reduction when failure starts propagating.
\begin{gather}\label{eq: load shed loss}
    R(p, \Delta w) = R^G(p, \Delta w) +  R^L(p, \Delta w),\\
    R^G(p, \Delta w) = G(p')-G(p)\\
    R^L(p, \Delta w) = L(p')-L(p),
\end{gather}
$R^G(p, \Delta w)$ and $R^L(p, \Delta w)$ correspond to grid-centric and consumer-centric resilience, respectively.
\subsection{Corrective Action Analysis} \label{Section 4.4: Results}
Our experiment finds that, under certain net loading levels, full service is impossible even without contingencies, as our solver fails to converge. The problem arises under scenarios at high loading levels or non-uniform shedding  priorities. In \textit{Experiment 2}, DC OPF and AC OPF fail to converge for loading levels greater than $1.3 \times$ and $1.0\times$ the default loading level, respectively. This signifies the necessity for smart scheduling. The rest of this section presents highlights of the analysis where initialization succeeds.

\begin{figure}[hbtp]
\centering
\includegraphics[width=0.3\textwidth]{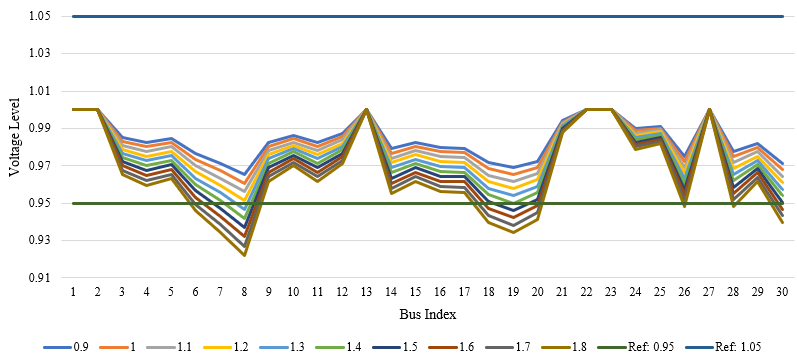}
\captionsetup{font={footnotesize}}\caption{Initial bus voltages solved with AC power flow.} \label{fig: ACPF}
\end{figure}

We find that PF solutions are frequently not physically implementable. This can be observed in the initial voltages under AC PF in \cref{fig: ACPF}. Bus voltages drops significantly as loading increases, falling outside the $(0.95, 1.05)$ constraint.

In all three experiments, AC models uncover more link failures and greater load shed than their DC model counterparts. In particular, in \textit{Experiment 1}, losses on link failure is only slightly greater in AC than DC models, but this difference is much greater in \textit{Experiment 2}. The high levels of losses from AC solutions renders the DC approximation insufficient to study failure cascades, as it underestimates the severity of contingencies. Findings about the load shed losses present an especially optimistic outlook. In \textit{Experiment 1 \& 2}, AC models render much higher load shed than DC simulations. However, in \textit{Experiment 3}, when cost-based flexible load shed is implemented, this load shed cost is no longer so significant. As AC simulation results are physically implementable, this result shows that we can serve close to full demand without causing congestion with optimal re-dispatch for both generation and load. This is especially promising in practice.

\begin{figure}[hbtp]
\begin{minipage}{0.21\textwidth}
\includegraphics[width=1\textwidth]{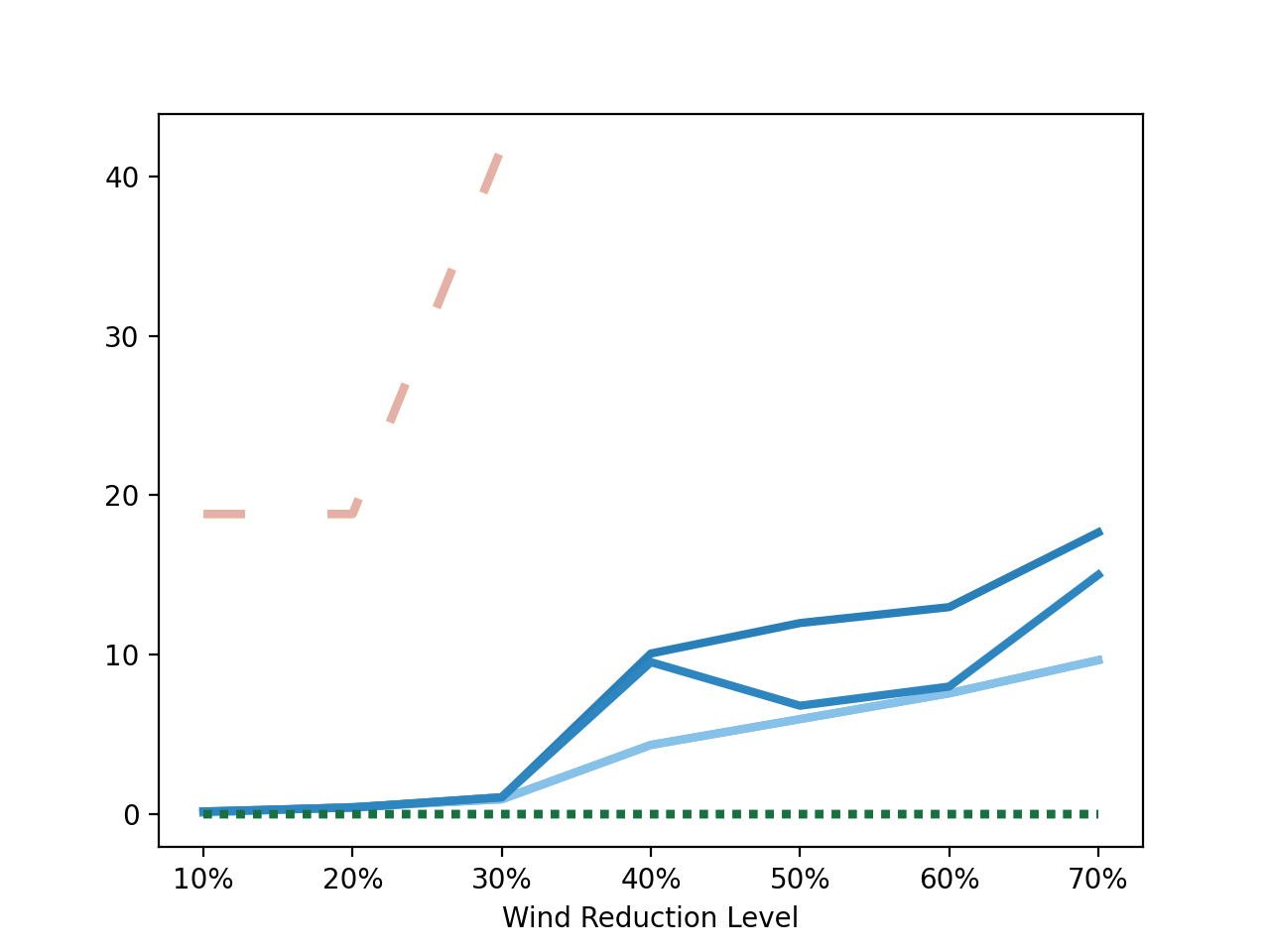}
\captionsetup{font={footnotesize}}\caption{Grid-Centric.} 
\label{fig: RG}
\end{minipage}\hfill
\begin{minipage}{0.21\textwidth}
\includegraphics[width=1\textwidth]{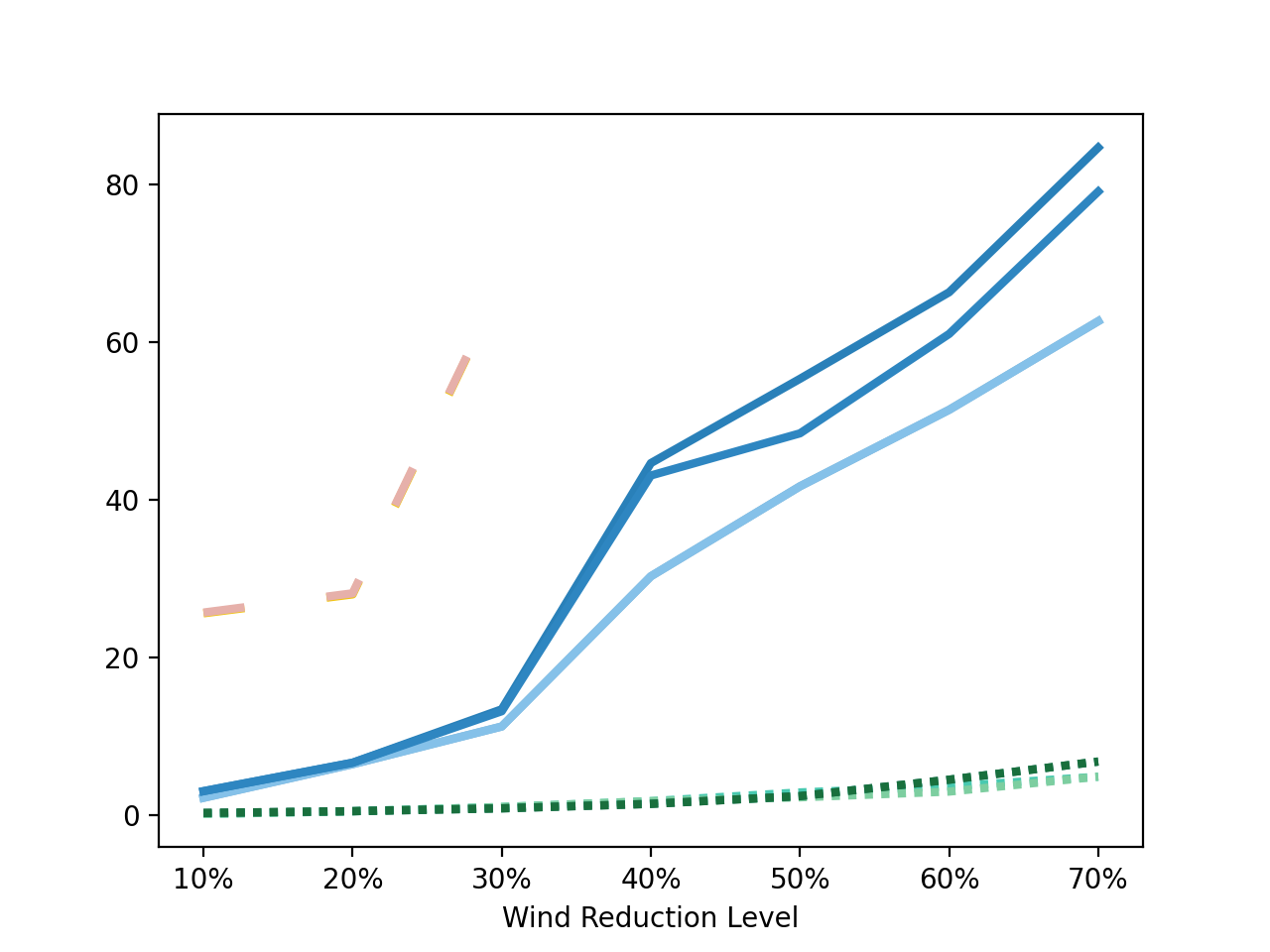}
\captionsetup{font={footnotesize}}\caption{Consumer-Centric.} 
\label{fig: RL}
\end{minipage}\hfill
\begin{minipage}{0.06\textwidth}
\includegraphics[width=1\textwidth]{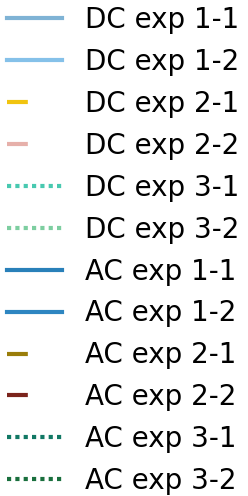}
\label{fig: legends}
\end{minipage}
\captionsetup{font={footnotesize}}\caption{Resilience Impact.}
\label{fig: resilience}
\end{figure}

To evaluate the resilience impact $R(p, \Delta w),$ it suffices to know $G(p), L(p), G(p'), L(p').$ We find that DC models significantly underestimate the impact of wind reduction for both resilience measures. We present the analysis of one particular scenario --- when the initial net load (load - wind power) is the system load, and examine the effects of different corrective actions upon wind power reduction up to $70\%$ of the system load. Our experiments found that, as a result of non-convergence, blackout happens when wind reduction is $\geq 40\%$ under DC models \textit{Experiment 2}, and any level of wind reduction will cause blackout under AC models. \cref{fig: resilience} presents the grid-centric ($R^G$) and consumer-centric ($R^L$) impact. For both $R^G$ and $R^L,$ all models find the impact to increase drastically at higher levels of wind reduction. Smart load re-dispatch (as in \textit{Exp. 3}) minimizes the impact, reducing it to about one-tenth of the impact when no action is taken (as in \textit{Exp. 1}) under both DC and AC models. This result underlines that, without proactive planning to prevent blackout, wind penetration is risky, as unexpected wind reduction of as little as $10\%$ of the base load can cause large-scale congestion, but the risk can be significantly reduced with smart rescheduling, which ensures near full service and avoids congestion altogether.

\subsection{System-Wide Structures} \label{section 4.5: structures}
\begin{figure}[hbtp]
\begin{minipage}{0.15\textwidth}
\includegraphics[width=1\textwidth]{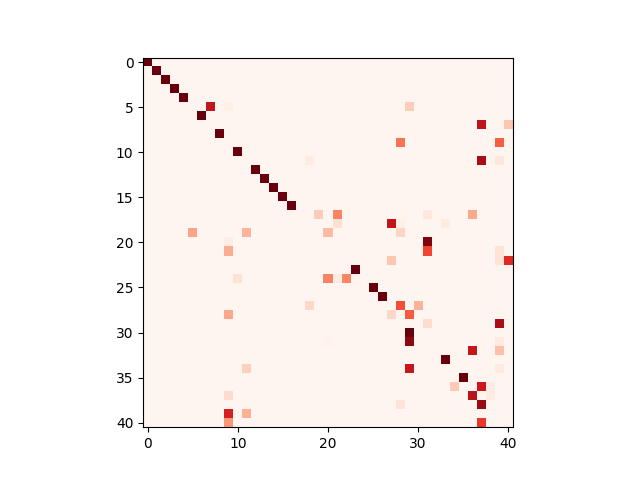}
\captionsetup{font={footnotesize}}\caption{$D$ matrix for DC PF, $1.6\times$ loading} 
\label{fig: D exp1-DC}
\end{minipage}\hfill
\begin{minipage}{0.15\textwidth}
\includegraphics[width=1\textwidth]{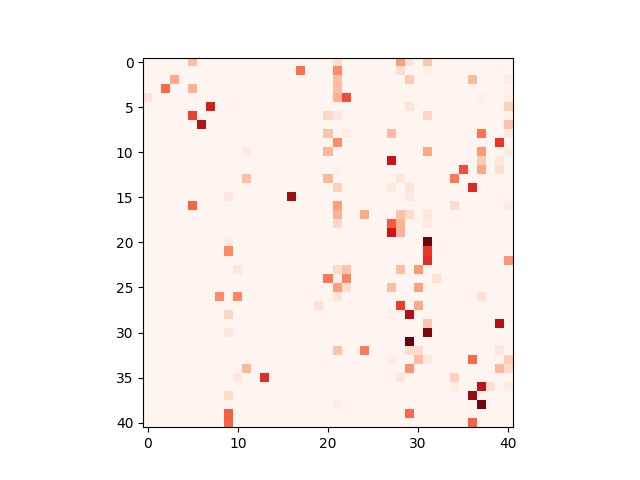}
\captionsetup{font={footnotesize}}\caption{$D$ matrix for AC PF, $1.6\times$ loading} 
\label{fig: D exp1-AC}
\end{minipage} \hfill
\begin{minipage}{0.15\textwidth}
\includegraphics[width=1\textwidth]{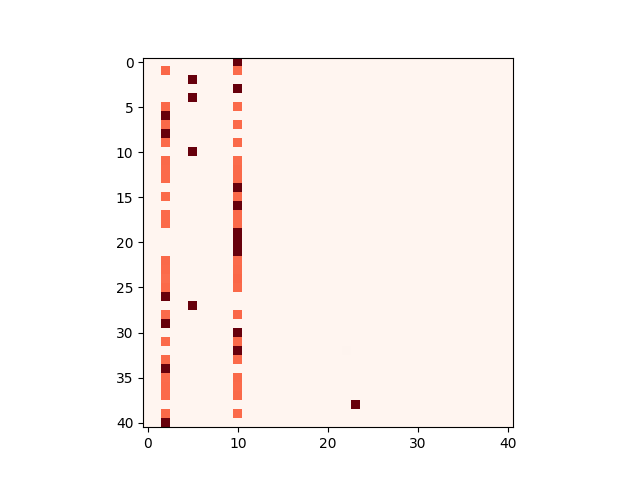}
\captionsetup{font={footnotesize}}\caption{$D$ matrix for DCOPF, $1\times$ loading} 
\label{fig: D exp2-DC}
\end{minipage} \\
\begin{minipage}{0.15\textwidth}
\includegraphics[width=1\textwidth]{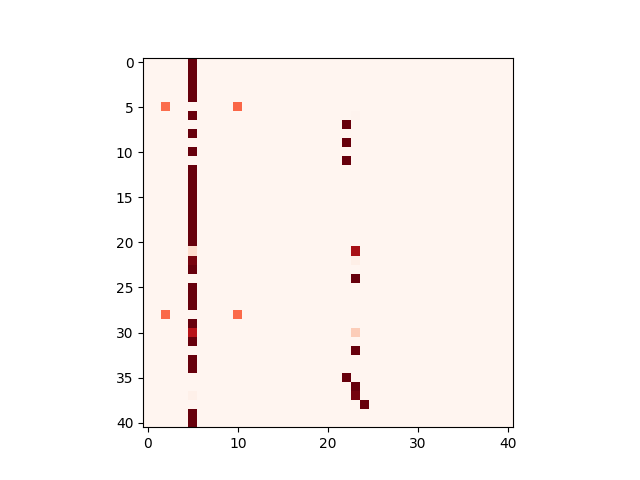}
\captionsetup{font={footnotesize}}\caption{$D$ matrix for ACOPF, $1\times$ loading} 
\label{fig: D exp2-AC}
\end{minipage}\hfill
\begin{minipage}{0.15\textwidth}
\includegraphics[width=1\textwidth]{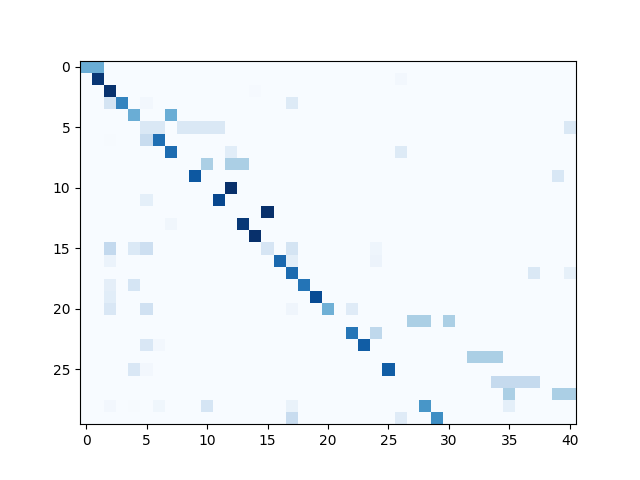}
\captionsetup{font={footnotesize}}\caption{$E$ matrix for DC PF, $1.6\times$ loading} 
\label{fig: E exp1-DC}
\end{minipage}\hfill
\begin{minipage}{0.15\textwidth}
\includegraphics[width=1\textwidth]{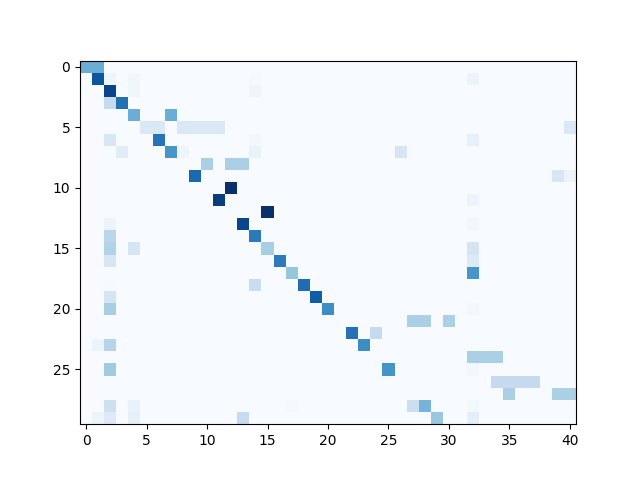}
\captionsetup{font={footnotesize}}\caption{$E$ matrix for AC PF, $1.6\times$ loading} 
\label{fig: E exp1-AC}
\end{minipage} 
\captionsetup{font={footnotesize}}\caption{$D, E$ matrix structures.}
\label{fig:DE}
\end{figure}

A few interesting system-wide influence structures arise from our $D$ and $E$ matrices. Heatmaps of selected scenarios for $D$ and $E$ matrices are shown in \cref{fig:DE}, where darker colors denote higher influence levels.  The $E$ matrices display a sparse structure. When no corrective actions are taken, the $D$ matrix has a sparse structure under DC models and linear structure under AC models (\cref{fig: D exp1-DC}, \cref{fig: D exp1-AC}). However, when corrective actions are taken, the $D$ matrices display a linear structure, where the pair-wise influence values are high in particular columns (\cref{fig: D exp2-DC}, \cref{fig: D exp2-AC}). This suggests a uni-directional, strong influence from one to many other links. 

This linear structure has significant practical value. Identifying the high-influence links in $D$ can be extremely informative to the operators: when critical links fail, there is higher value to execute scheduling according to the scheme in \textit{Experiment 3} to preserve infrastructure integrity.
\section{Prediction Accuracy} \label{Section 5: Prediction Accuracy}
Both link failure and load shed prediction reach high prediction accuracy, $>90\%$ for most cases and $>80\%$ for all. There is no significant difference between the training and testing sets, which verifies that our model does not overfit. We find no significant difference between the DC and AC models. The IM framework provide significant reduction in computational time, especially for the AC models, which requires solving nonlinear equations. While flow-based solutions become more inefficient at higher loading level, the computational cost is identical for all cases under our method.
To demonstrate that our proposed framework scheme captures the failure cascade features, we compare it to two prediction methods that do not depend on the IM: the uniform, deterministic prediction and the randomized prediction. The IM methods gave better performance than both random and uniform predictions with no significant difference across different loading levels as shown by the mean error rates in \cref{Table 3.1: link fail pred} and \cref{Table 3.2: load shed pred}.

\begin{table}[hbtp]
\begin{minipage}{0.22\textwidth}
\begin{tabular}{l|lll}
     & IM    & Rand. & Unif. \\ \hline
exp1 & 0.038 & 0.188  & 0.109   \\
exp2 & 0.019 & 0.093  & 0.049   \\
exp3 & 0.000 & 0.094  & 0.049  
\end{tabular}\captionsetup{font={footnotesize}}\caption{Link Failure Prediction Error Rates.} \label{Table 3.1: link fail pred}
\end{minipage}\hfill
\begin{minipage}{0.22\textwidth}
\begin{tabular}{l|lll}
     & IM    & Rand. & Unif. \\ \hline
exp1 & 0.214 & 0.318  & 0.255   \\
exp2 & 0.043 & 0.082  & 0.043   \\
exp3 & 0.014 & 0.026  & 0.014  
\end{tabular}
\captionsetup{font={footnotesize}}\caption{Load Shed Prediction Error Rates.} \label{Table 3.2: load shed pred}
\end{minipage} 
\end{table}

\section{Tests in Large-Scale IEEE 300 Bus System} \label{Section 6: Large System}
We verify that our methodology is accurate and scalable by experimenting on the IEEE300 system\cite{MATPOWER}.
In this system, smart scheduling in \textit{Experiment 3} can reduce the cost of load shed by as much as $90\%$ at all loading levels, while completely avoiding all link failures.
Prediction with the IM model yields even higher accuracy, with $>95\%$ for most cases, for both link failure and load shed, with the exception of load shed prediction in \textit{Experiment 1}, which also produced $>75\%$ accuracy (the latter requires further assessment). 
The influence model has significant computation cost advantage. \cref{Table: LS time cost} shows an example at the default loading level on 1000 samples. The influence model can predict link failures and load shed within 1/10 of the time required for running PF or OPF \footnote{Tested with MATLAB R2022a on Intel(R)
Core(TM) i5-1135G7 CPU@2.40GHz Processor with 8GB
installed memory.}. Further studies are needed to assess the economic advantage of our adaptive approach of rescheduling generation upon contingencies.  
 \begin{table}[h]
\centering
\begin{tabular}{l|lll}
     & Simulation (s)    & Training (s) & Prediction (s) \\ \hline
exp1 & 169.77 & 611.50  & 15.40  \\
exp2 & 183.35 & 305.63  & 10.05   \\
exp3 & 246.23 & 332.68   & 6.76  
\end{tabular}
\captionsetup{font={footnotesize}}\caption{Time Cost at Default Loading Level.} \label{Table: LS time cost}
\end{table}

\section{Conclusion: Advisory Tool for Operators} \label{Section 7: Advisory tool for operators}
Based on our study of (1) the impact of wind reduction and effects of corrective actions during a failure cascade and (2) the resilience impact measures and risks of wind reduction, we propose an advisory tool for operators to assess contingency criticality, predict losses, and strategize for loss minimization  with smart rescheduling. The tests have shown promising robust results. The advisory tool comprises:
\subsection{Wind Reduction Risk Assessment} \label{Section 7.1: risk assessment}
The IM helps us determine the most critical links and initial contingencies, determined by a combination of criticality values as well as expected $G(p), L(p),$ and $R(p, \cdot)$ where profile $p$ embeds the initial contingencies. Grid and consumer-centric criticality values are computed with equations as follows.
\begin{gather}
    C^D(j) = \sum_{i=1, \dots, N_{br}} d_{ij}(A^{11}_{ji}-A^{01}_{ji}),\\
    C^E(j) = \sum_{i=1, \dots, N} e_{ij}(B^{11}_{ji}-B^{01}_{ji}),
\end{gather}
where $j$ is the link index, and $i$ enumerates over all links for $C^D(j)$ or all buses for $C^E(j)$.
With this information, system operators can identify the impact of sudden wind reduction when there is a pre-existing link failure in the network. As greater values of the resilience impact signify greater cost, this tool informs operators of upcoming risks.
\subsection{Cascade Management} \label{Section 7.2: cascade management}
The failure cascade and load shed prediction given by our models can inform operators the best course of action for loss minimization. Operators may also use this tool to predict the impact of wind reduction and deploy load reduction.

\end{document}